\documentstyle[12pt,axodraw]{article}

\catcode`@=12
\begin{document}
\thispagestyle{empty}
\begin{flushright} 
UCRHEP-T312\\ 
July 2001\
\end{flushright}
\vspace{0.5in}
\begin{center}
{\LARGE	\bf Light Unstable Sterile Neutrino\\}
\vspace{1.5in}
{\bf Ernest Ma$^a$ and G. Rajasekaran$^b$\\}
\vspace{0.2in}
{\sl $^a$ Physics Department, University of California, Riverside, 
California 92521\\}
\vspace{0.1in}
{\sl $^b$ Institute of Mathematical Sciences, Chennai (Madras) 600113, India\\}
\vspace{1.5in}
\end{center}
\begin{abstract}\
The three massless active (doublet) neutrinos may mix with two heavy and one 
\underline {light} sterile (singlet) neutrinos so that the induced masses 
and mixings among the former are able to explain the present data on 
atmospheric and solar neutrino oscillations.  If the LSND result is also to 
be explained, one active neutrino mass eigenstate must mix with the light 
sterile neutrino.  A specific model is proposed with the spontaneous and soft 
explicit breaking of a new global $U(1)_S$ symmetry so that a sterile 
neutrino will decay into an active antineutrino and a nearly massless 
pseudo-Majoron.
\end{abstract}
\newpage
\baselineskip 24pt

Present experimental data \cite{atm,solar,lsnd} indicate that neutrinos 
oscillate.  Hence they should have small nonzero masses and mix with one 
another.  This may be achieved without additional fermions beyond those of 
the minimal standard model by a heavy Higgs triplet \cite{masa,marasa}.  On 
the other hand, most theoretical approaches assume the addition of 3 singlet 
neutral fermions (usually considered as right-handed neutrinos $N_R$). 
In that case, a Dirac mass $m_D$ linking the left-handed doublet neutrinos 
$\nu_L$ with $N_R$ as well as a Majorana mass $M$ for $N_R$ are allowed, 
thus yielding the famous mass matrix
\begin{equation}
{\cal M}_{\nu N} = \left( \begin{array} {c@{\quad}c} 0 & m_D \\ m_D & M 
\end{array} \right).
\end{equation}
At this point, one may impose the conservation of lepton number as an 
additive global symmetry, i.e. $U(1)_L$, so that $M=0$; but then $m_D$ 
would have to be extremely small, which is considered rather unnatural.  
The conventional solution of this problem is to not consider $U(1)_L$ 
at all so that $M$ is naturally very large and since $m_D$ cannot be larger 
than the electroweak breaking scale $v = (2 \sqrt 2 G_F)^{-1/2} = 174$ GeV, a 
small mass  $m_\nu = m_D^2/M$ is obtained \cite{seesaw}.  This of course 
requires $M$ to be many orders of magnitude greater than $v$ and renders it 
totally undetectable experimentally.  Recently, it has been pointed out 
\cite{ma01} that if $m_D$ comes from a different Higgs doublet with a 
suppressed vacuum expectation value (VEV), then $M$ may in fact be only 
a few TeV or less and become observable at future colliders.

In this note we consider the case where both $m_D$ and $M$ are small for 
\underline {one} (call it $S$) of the three singlets, but 
$m_D$ is still less than $M$ by perhaps an order of magnitude.  This is 
in contrast to the pseudo-Dirac scenario \cite{koli}, i.e. $M << m_D$, in 
which case neutrino oscillations would be maximal between active and sterile 
species, in disfavor with the most recent data \cite{atm,solar}.  Before 
discussing the theoretical reasons for $m_D$ and $M$ to be small, consider 
first the phenomenology of such a possibility.  The 3 active neutrinos 
$\nu_e$, $\nu_\mu$, $\nu_\tau$ are now each a linear combination of 4 
light neutrino mass eigenstates.  With $m_D$ less than $M$ by an order of 
magnitude, the mixing of $S$ with $\nu$ is still small; hence the 
presumably large mixings among the 3 active neutrinos themselves are 
sufficient to explain the atmospheric \cite{atm} and solar \cite{solar} 
neutrino data.  This leaves the LSND data \cite{lsnd} to be explained by 
having a neutrino mass eigenstate which is mostly $S$ but with small 
amounts of $\nu_e$ and $\nu_\mu$.

In addition to the one light $S$ and the two heavy 
$N$'s, we supplement the particle content of the standard model with a scalar 
singlet $\chi^0$ and an extra scalar doublet $\eta = (\eta^+, \eta^0)$, 
together with a new global $U(1)_S$ symmetry such that $(S, \chi^0, \eta)$ 
have charges $(1,-2,-1)$ respectively.  The relevant terms of the Lagrangian 
involving these fields are then given by
\begin{equation}
h \chi^0 S S + f_i S (\nu_i \eta^0 - l_i \eta^+) + h.c.
\end{equation}

Using the canonical seesaw mechanism \cite{seesaw} with the two heavy $N$'s, 
we obtain two massive neutrino eigenstates in the conventional way.  The 
original $6 \times 6$ neutrino mass matrix is reduced to a $4 \times 4$ 
matrix spanning $(\nu_1, \nu_2, \nu_3, S)$.  Its most general form is given by
\begin{equation}
{\cal M}_{\nu S} = \left[ \begin{array} {c@{\quad}c@{\quad}c@{\quad}c} 
0 & 0 & 0 & \mu_1 \\ 0 & m'_2 & 0 & \mu_2 \\ 0 & 0 & m'_3 & \mu_3 \\ 
\mu_1 & \mu_2 & \mu_3 & M \end{array} \right],
\end{equation}
where $M = 2 h \langle \chi^0 \rangle$ and $\mu_i = f_i \langle \eta^0 
\rangle$.

To obtain $\langle \eta^0 \rangle \sim 0.1$ eV, consider the part of the 
Higgs potential involving $\eta$, i.e.
\begin{equation}
V_\eta = m_\eta^2 \eta^\dagger \eta + {1 \over 2} \lambda_1 (\eta^\dagger 
\eta)^2 + \lambda_3 (\eta^\dagger \eta)(\Phi^\dagger \Phi) + \lambda_4 
(\eta^\dagger \Phi)(\Phi^\dagger \eta) + [\mu_0^2 \eta^\dagger \Phi + h.c.],
\end{equation}
where $\Phi$ is the usual standard-model Higgs doublet and the $\mu_0^2$ 
term breaks $U(1)_S$ softly.  The equation of constraint for $\langle \eta^0 
\rangle = u$ is then given by
\begin{equation}
u[m_\eta^2 + \lambda_1 u^2 + (\lambda_3 + \lambda_4) v^2] + \mu_0^2 v = 0,
\end{equation}
where $v = \langle \phi^0 \rangle$.  For $m_\eta^2 > 0$ and large, we then 
have
\begin{equation}
u \simeq - {\mu_0^2 v \over m_\eta^2}.
\end{equation}
Let $m_\eta \sim 1$ TeV and $\mu_0 \sim 1$ MeV, we obtain $u \sim 0.1$ eV 
as desired.

To obtain $z = \langle \chi^0 \rangle \sim 1$ eV, we use the $shining$ 
mechanism \cite{shine} of large extra dimensions, where $\chi^0$ is assumed 
to exist in the bulk and its VEV on our brane is suppressed because of its 
distance from the source brane of $U(1)_S$ breaking.  For consistency, the 
$\chi^0 S S$ interaction is replaced by $z \exp (i\sqrt 2 \varphi/z) S S $.  
This has been explained fully in a previous paper \cite{marajsa}.  The 
important difference here is that $U(1)_S$ is also broken explicitly so 
that the would-be massless Goldstone boson $\varphi$, i.e. the Majoron 
\cite{majoron,valle}, is not strictly massless.  On the other hand, its 
mass may still be very small.  We may call it a pseudo-Majoron.

Returning to Eq.~(3), we assume for definiteness a bimaximal pattern of 
mixing among the active neutrinos, i.e.
\begin{equation}
\left[ \begin{array} {c} \nu_1 \\ \nu'_2 \\ \nu_3 \end{array} \right] = 
\left[ \begin{array} {c@{\quad}c@{\quad}c} 1/\sqrt 2 & 1/2 & 1/2 \\ 
-1/\sqrt 2 & 1/2 & 1/2 \\ 0 & -1/\sqrt 2 & 1/\sqrt 2 \end{array} \right] 
\left[ \begin{array} {c} \nu_e \\ \nu_\mu \\ \nu_\tau \end{array} \right],
\end{equation}
together with the {\it ansatz} that $\mu_1$ and $\mu_3$ are negligible. 
In that case, only $\nu'_2$ mixes significantly with $S$.  The eigenstates 
are thus $\nu'_2 \cos \theta + S \sin \theta$ with mass $m'_2 - \mu_2^2/M \sim 
0.007$ eV and $S \cos \theta - \nu'_2 \sin \theta$ with mass $M \sim$ few eV, 
where $\sin \theta \simeq -\mu_2/M$.  Hence the latter decays into the 
conjugate of the former and the pseudo-Majoron with coupling $2 \sqrt 2 h 
\sin \theta \cos \theta$.  [If all $\mu_i$'s were of the same order of 
magnitude, the present observed neutrino oscillations cannot be explained, 
unless the 3 active neutrinos are almost degenerate in mass, requiring thus 
a high degree of unnatural fine tuning of parameters.  Also, the nonzero 
overlap with $S$ would make $\nu_3$ and $\nu_2$ decay into $\nu_1$.]

The $\nu_\mu \to \nu_e$ probability in the LSND experiment is given by 
\cite{marajst}
\begin{equation}
P_{\mu e} = {s^4 \over 8} \left( 1 + x^2 - 2x \cos {M^2 L \over 2E} \right) 
\sim 10^{-3},
\end{equation}
where $s = \sin \theta$ and $x = \exp (-M \Gamma L/2E)$ is the decay  
factor. [In the usual case of a stable sterile neutrino, $\Gamma = 0$ so 
$x=1$.] The decay rate $\Gamma$ is easily calculated to be
\begin{equation}
\Gamma = {h^2 s^2 c^2 M \over 2 \pi} \simeq 0.18 M \left( {h^2 \over 4 \pi} 
\right) \left( {s^2 \over 0.1} \right) \left( {c^2 \over 0.9} \right),
\end{equation}
which is of the right order of magnitude for it to be significant 
\cite{marajst} in affecting the interpretation of the LSND data in terms 
of \underline {both} oscillation and decay.

The $4 \times 4$ neutrino mixing matrix is now given by
\begin{equation}
\left[ \begin{array} {c} \nu_1 \\ \nu_2 \\ \nu_3 \\ \nu_4 \end{array} \right] 
= \left[ \begin{array} {c@{\quad}c@{\quad}c@{\quad}c} 1/\sqrt 2 & 1/2 & 1/2 
& 0 \\ -c/\sqrt 2 & c/2 & c/2 & s \\ 0 & -1/\sqrt 2 & 1/\sqrt 2 
& 0 \\ s/\sqrt 2 & -s/2 & -s/2 & c \end{array} \right] \left[ \begin{array} 
{c} \nu_e \\ \nu_\mu \\ \nu_\tau \\ S \end{array} \right],
\end{equation}
with $m_1 \simeq 0$, $m_2 \simeq m'_2 - \mu_2^2/M \simeq 0.007$ eV, $m_3 
\simeq m'_3 \simeq 0.05$ eV, and $m_4 \simeq M \sim$ few eV.  The 
phenomenology of this scheme for atmospheric and solar neutrino oscillations 
has been fully described previously \cite{marajst}.  We emphasize here the 
most important prediction of this model, i.e. the decay
\begin{equation}
\nu_4 \to \bar \nu_2 + \zeta,
\end{equation}
where $\zeta$ is the pseudo-Majoron. 
Since $\nu_e$ from the Sun has a $\nu_4$ component, it will decay into 
$\bar \nu_2$ on its way to the Earth.  The latter will be observed as 
$\bar \nu_e$ in detectors such as BOREXINO and perhaps SNO.  The advantage 
of having $\nu_4$ decay is to evade the indirect constraint from the 
CDHSW experiment \cite{cdhsw} on the LSND allowed parameter space for 
neutrino oscillations \cite{marajst}.  Without decay, the $(3+1)$ scheme 
of neutrino masses may be disfavored \cite{3+1}.  Note also that in our 
model, the pseudo-Majoron does not couple to the active neutrinos, otherwise 
there would be significant bounds on the corresponding coupling strengths 
\cite{bounds}.

The effective number of neutrinos $N_\nu$ for successful nucleosynthesis 
\cite{bbn} is probably not greater than 4.  In our scenario, it appears that 
$N_\nu = 4 + (8/7)$, counting as well $S$ and $\chi^0$.  However, 
these two fields decouple from the standard-model particles at the scale 
$M_\eta$ which we take to be 1 TeV.  This means that whereas 
$\nu_{e,\mu,\tau}$ are heated by the subsequent annihilations of 
nonrelativistic particles, $S$ and $\chi^0$ are not \cite{oss}.  Thus the 
number densities of the latter are greatly suppressed at the time of 
nucleosynthesis in the early Universe and $N_\nu < 4$ is easily obtained 
\cite{eens}.

In conclusion, we have constructed a specific model in this short note in 
the framework of 3 active (doublet) and 3 sterile (singlet) neutrinos.  
Two of the latter are heavy, providing small seesaw masses for two active 
neutrinos.  The third sterile neutrino is light and mixes with one of the 
massive active neutrinos.  Together they allow all neutrino-oscillation data 
to be explained in a hierarchical pattern of neutrino masses.  The light 
sterile neutrino is associated with a new global $U(1)_S$ symmetry which is 
spontaneously and softly broken, so that it decays into an active antineutrino 
and a nearly massless pseudo-Majoron.

\vskip 0.3in
This work was supported in part by the U.~S.~Department of Energy
under Grant No.~DE-FG03-94ER40837.  G.R. also thanks the UCR Physics 
Department for hospitality.

\newpage
\bibliographystyle{unsrt}

\end{document}